\let\includefigures=\iftrue   
\input harvmac 


\input epsf 

\newcount\figno 
\figno=0 
\def\fig#1#2#3{ 
\par\begingroup\parindent=0pt\leftskip=1cm\rightskip=1cm\parindent=0pt 
\baselineskip=11pt 
\global\advance\figno by 1 
\midinsert 
\epsfxsize=#3 
\centerline{\epsfbox{#2}} 
\vskip 12pt 
{\bf Figure \the\figno:} #1\par 
\endinsert\endgroup\par } 
\def\figlabel#1{\xdef#1{\the\figno}} 


\noblackbox 
\def\IZ{\relax\ifmmode\mathchoice 
{\hbox{\cmss Z\kern-.4em Z}}{\hbox{\cmss Z\kern-.4em Z}} 
{\lower.9pt\hbox{\cmsss Z\kern-.4em Z}} {\lower1.2pt\hbox{\cmsss 
Z\kern-.4em Z}}\else{\cmss Z\kern-.4em Z}\fi}

\def\p{\partial}

\font\cmss=cmss10 \font\cmsss=cmss10 at 7pt 
\def\IR{\relax{\rm I\kern-.18em R}} 
 
\def\frac#1#2{{#1 \over #2}}


\def\journal#1&#2(#3){\unskip, \sl #1\ \bf #2 \rm(19#3) }   
\def\andjournal#1&#2(#3){\sl #1~\bf #2 \rm (19#3) }

%
%

%
\catcode`\@=11   
\def\slash#1{\mathord{\mathpalette\c@ncel{#1}}}   
\overfullrule=0pt

\def\underrel#1\over#2{\mathrel{\mathop{\kern\z@#1}\limits_{#2}}}

\def\in{\infty}
\def\go{\rightarrow}
\def\co{coordinates}

\catcode`\@=12   
   
   
%


\def\myTitle#1#2{\nopagenumbers\abstractfont\hsize=\hstitle\rightline{#1}%
\vskip 0.5in\centerline{\titlefont #2}\abstractfont\vskip .5in\pageno=0} 

\myTitle{\vbox{\baselineskip12pt\hbox{} 
\hbox{RI-03-03} 
\hbox{ITFA-2003-12}
}} {\vbox{ 
        \centerline{Penrose limit and DLCQ of string theory} 
        \medskip 
        }} 
\centerline{Assaf Shomer\foot{E-mail : {\tt shomer@cc.huji.ac.il}$\quad;\quad $ 
{\tt shomer@science.uva.nl.}}} 
\medskip 
\centerline{Racah Institute of Physics, The Hebrew University, 
Jerusalem 91904, Israel} 
\medskip 
\centerline{and}
\medskip
\centerline{Institute of Theoretical Physics, University of Amsterdam}
 \centerline{Valckenierstraat 65, 1018 XE Amsterdam, The Netherlands.} 
\bigskip 
\bigskip 
\noindent 

We argue that the Penrose limit of a general string background is a 
generalization of the Seiberg-Sen limit describing M(atrix) theory as the DLCQ 
of M theory in flat space.
The BMN theory of type IIB strings on the maximally supersymmetric pp-wave 
background is understood as the exact analogue of the BFSS M(atrix) theory, 
namely, a DLCQ of IIB string theory on $AdS_5 \times S^5$ in the limit of 
infinite longitudinal momentum. This point of view is used to explain some 
features of the BMN duality.

\Date{March 2003} 

\baselineskip=16pt

\lref\witads{E.~Witten,``Anti-de Sitter space and holography,'' 
Adv.\ Theor.\ Math.\ Phys.\  {\bf 2} (1998) 253 ; hep-th/9802150.} 

\lref\juan{J.~Maldacena, 
``The large $N$ limit of superconformal field theories and supergravity,'' 
Adv.\ Theor.\ Math.\ Phys.\  {\bf 2} (1998) 231 ; 
Int.\ J.\ Theor.\ Phys.\  {\bf 38} (1998) 1113 ; hep-th/9711200.}

\lref\adsrev{O.~Aharony, S.~S.~Gubser, J.~Maldacena, H.~Ooguri and Y.~Oz, 
``Large N field theories, string theory and gravity,'' 
Phys.\ Rept.\  {\bf 323} (2000) 183 ; hep-th/9905111.
E.~D'Hoker and D.~Z.~Freedman,
``Supersymmetric gauge theories and the AdS/CFT correspondence,'' ;
hep-th/0201253.}

\lref\gkp{S.~S.~Gubser, I.~R.~Klebanov and A.~M.~Polyakov,
``Gauge theory correlators from non-critical string theory,''
Phys.\ Lett.\ B {\bf 428} (1998) 105
; hep-th/9802109.}

\lref\dvv{R.~Dijkgraaf, E.~Verlinde and H.~Verlinde,
``Matrix string theory,'' Nucl.\ Phys.\ B {\bf 500} (1997) 43 ; hep-th/9703030.}

\lref\bs{T.~Banks and N.~Seiberg,``Strings from matrices,'' Nucl.\ Phys.\ B {\bf 
497} (1997) 41 ; hep-th/9702187.}

\lref\mrv{S.~Mukhi, M.~Rangamani and E.~Verlinde,
``Strings from quivers, membranes from moose,''
JHEP {\bf 0205} (2002) 023 ; hep-th/0204147.}

\lref\guven{R.~Gueven,
``Plane wave limits and T-duality,''
Phys.\ Lett.\ B {\bf 482} (2000) 255 ; hep-th/0005061.}

\lref\bmn{D.~Berenstein, J.~M.~Maldacena and H.~Nastase,
``Strings in flat space and pp waves from N = 4 super Yang Mills,''
JHEP {\bf 0204} (2002) 013 ; hep-th/0202021.}

\lref\bdhio{M.~Bertolini, J.~de Boer, T.~Harmark, E.~Imeroni and N.~A.~Obers,
``Gauge theory description of compactified pp-waves,''
JHEP {\bf 0301} (2003) 016 ; hep-th/0209201.}

\lref\bfss{T.~Banks, W.~Fischler, S.~H.~Shenker and L.~Susskind,
``M theory as a matrix model: A conjecture,''
Phys.\ Rev.\ D {\bf 55} (1997) 5112 ; hep-th/9610043.}

\lref\susn{L.~Susskind,
``Another conjecture about M(atrix) theory,''; hep-th/9704080.}

\lref\seiberg{N.~Seiberg,
``Why is the matrix model correct?,''
Phys.\ Rev.\ Lett.\  {\bf 79} (1997) 3577 ; hep-th/9710009.}

\lref\sen{A.~Sen,
``D0 branes on T(n) and matrix theory,''
Adv.\ Theor.\ Math.\ Phys.\  {\bf 2} (1998) 51 ; hep-th/9709220.}

\lref\bigsus{D.~Bigatti and L.~Susskind,
``Review of matrix theory,'' ; hep-th/9712072.}

\lref\banks{T.~Banks,
``Matrix theory,''
Nucl.\ Phys.\ Proc.\ Suppl.\  {\bf 67} (1998) 180
; hep-th/9710231.} 

\lref\joe{J.~Polchinski,
``M-theory and the light cone,''
Prog.\ Theor.\ Phys.\ Suppl.\  {\bf 134} (1999) 158
; hep-th/9903165.}

\lref\guijosa{A.~Guijosa,
compactification 
radius?,'' 
Nucl.\ Phys.\ B {\bf 533} (1998) 406 ; hep-th/9804034.}

\lref\bseven{N.~R. Constable { et al.}, 
``Pp-wave string interactions from perturbative {Yang--Mills} theory,'' 
{JHEP} {\bf 07} (2002) 017 ; 
hep-th/0205089.}

\lref\plefka{C.~Kristjansen, J.~Plefka, G.~W.~Semenoff and M.~Staudacher,
``A new double-scaling limit of N = 4 super Yang-Mills theory and PP-wave 
strings,'' 
Nucl.\ Phys.\ B {\bf 643}, 3 (2002) ; hep-th/0205033.}

\lref\motl{L.~Motl, ``Proposals on nonperturbative superstring interactions'' ; 
hep-th/9701025.}

\lref\wati{W.~Taylor,
``M(atrix) theory: Matrix quantum mechanics as a fundamental theory,''
Rev.\ Mod.\ Phys.\  {\bf 73} (2001) 419
; hep-th/0101126.}

\lref\imf{S.~Weinberg,
Phys.\ Rev.\  {\bf 150} (1966) 1313.
J.~B.~Kogut and L.~Susskind,
Phys.\ Rept.\  {\bf 8} (1973) 75.}

\lref\penrose{R.~Penrose, 
``Any space-time has a plane wave as a limit,''
in {Differential geometry and relativity}, Reidel, Dordrecht (1976) 271--275.}

\lref\weinberg{S.~Weinberg,
``The Quantum Theory Of Fields. Vol. 1: Foundations''}

\lref\nrst{J.~Gomis and H.~Ooguri,
``Non-relativistic closed string theory''
J.\ Math.\ Phys.\  {\bf 42} (2001) 3127
; hep-th/0009181. U.~H.~Danielsson, A.~Guijosa and M.~Kruczenski,
``IIA/B, wound and wrapped,'' 
JHEP {\bf 0010}, 020 (2000) ; hep-th/0009182.}

\lref\iw{E.~Inonu and E.~P.~Wigner,
Proc.\ Nat.\ Acad.\ Sci.\  {\bf 39} (1953) 510.}

\lref\gkptwo{S.~S.~Gubser, I.~R.~Klebanov and A.~M.~Polyakov,
Nucl.\ Phys.\ B {\bf 636} (2002) 99 ; hep-th/0204051.}

\lref\contjap{M.~Hatsuda, K.~Kamimura and M.~Sakaguchi,
Nucl.\ Phys.\ B {\bf 632} (2002) 114 ; hep-th/0202190.
M.~Hatsuda, K.~Kamimura and M.~Sakaguchi,
Nucl.\ Phys.\ B {\bf 637} (2002) 168 ; hep-th/0204002.}

\lref\ppw{M.~Blau, J.~Figueroa-O'Farrill, C.~Hull and G.~Papadopoulos,
``Penrose limits and maximal supersymmetry,''
Class.\ Quant.\ Grav.\  {\bf 19} (2002) L87 ; hep-th/0201081.
M.~Blau, J.~Figueroa-O'Farrill, C.~Hull and G.~Papadopoulos,
``A new maximally supersymmetric background of IIB superstring theory,''
JHEP {\bf 0201} (2002) 047
; hep-th/0110242. J.~Figueroa-O'Farrill and G.~Papadopoulos,
``Homogeneous fluxes, branes and a maximally supersymmetric solution of  
M-theory,''JHEP {\bf 0108} (2001) 036 ; hep-th/0105308.}

\lref\metsaev{R.~R. Metsaev and A.~A. Tseytlin, 
{Phys. Rev.} {\bf D65} 
(2002) 126004 ; hep-th/0202109. R.~R. Metsaev, 
{Nucl. Phys.} {\bf 
B625} (2002) 70--96 ; hep-th/0112044.}

\lref\tseytlin{J.~G.~Russo and A.~A.~Tseytlin,
JHEP {\bf 0209} (2002) 035 ; hep-th/0208114.
J.~G.~Russo and A.~A.~Tseytlin,
JHEP {\bf 0204} (2002) 021 ; hep-th/0202179.
R.~R.~Metsaev and A.~A.~Tseytlin,
Phys.\ Rev.\ D {\bf 65} (2002) 126004 ; hep-th/0202109. }

\lref\eliezer{K.~Sfetsos,
Phys.\ Lett.\ B {\bf 324} (1994) 335
; hep-th/9311010.
D.~I.~Olive, E.~Rabinovici and A.~Schwimmer,
Phys.\ Lett.\ B {\bf 321} (1994) 361
; hep-th/9311081.}

\lref\witsus{L.~Susskind and E.~Witten,
; hep-th/9805114.}

\lref\aki{F.~Antonuccio, A.~Hashimoto, O.~Lunin and S.~Pinsky,
JHEP {\bf 9907} (1999) 029  ; hep-th/9906087. I.~Chepelev,
Phys.\ Lett.\ B {\bf 453} (1999) 245
; hep-th/9901033.}

\lref\Helling{R.~Helling, J.~Plefka, M.~Serone and A.~Waldron,
``Three graviton scattering in M-theory,''
Nucl.\ Phys.\ B {\bf 559} (1999) 184
; hep-th/9905183.}

\newsec{Introduction.} 

One year ago Berenstein, Maldacena and Nastase (BMN) \bmn, building on recent 
work \refs{\ppw,\metsaev} showed that by taking a certain limit on both sides of 
the $AdS_5 \times S^5/SYM_4$ duality \juan\ one can find a sector of the $N=4$ 
SYM 
theory that describes, besides the supergravity modes, also the excited string 
states of type IIB string theory in a certain pp-wave 
background. String theory 
in that background is exactly solvable in lightcone gauge quantization 
\refs{\metsaev,\tseytlin} (see also \eliezer). 
BMN used this exact solution of type IIB 
string theory to predict the anomalous dimensions of 
operators in a certain sector of very large conformal dimensions and $R$ charge 
in the $N=4$ theory\foot{An alternative derivation of BMN's prediction using a 
somewhat different perspective was given in \gkptwo.}.    
The limit on the string side is called a ``Penrose limit" 
\penrose. The pp-wave background is the Penrose limit of $AdS_5 \times S^5$. 
It is a maximally supersymmetric solution to the type IIB 
supergravity equations of motion and an exact string background. 

However, the exact nature of this ``BMN duality" remains somewhat unclear. 
In the 
AdS/CFT correspondence \refs{\juan,\adsrev} the duality had a very precise and 
operational meaning 
\refs{\witads,\gkp}. One identified the generating function of correlators in 
the $N=4$ gauge theory with the partition function of string theory in $AdS_5 
\times S^5$ with appropriate vertex operator insertions on the boundary.
Type IIB strings on the maximally supersymmetric pp-wave are dual to a certain 
{\it sector} of the $N=4$ theory. This is a little disturbing since the one side 
seems like a well defined and self contained (string) theory while the other 
side is just a part of a (gauge) theory. Described loosely, it is not clear what 
we are equating on both sides. For example, the pp-wave background can be 
roughly described as having a 
confining harmonic potential which effectively reduces the theory to become 
$1+1$ dimensional \bseven. This makes it hard to give an $S$ matrix 
interpretation along the lines of \refs{\witads,\gkp}.  

In this paper we point out a close analogy between the 
Penrose limit and the Seiberg-Sen limit \refs{\seiberg,\sen} used in explaining 
M(atrix) theory. We interpret this in the following way. The Penrose limit of a 
string background should be viewed as a generalization of the Seiberg-Sen limit 
in flat space.
String/M theory in the Penrose limit of a space should be understood as a 
generalized DLCQ description of string/M theory in that space.
The BMN theory can be understood as a (generalized) DLCQ of type IIB strings on 
$AdS_5 \times S^5$, or equivalently, of $N=4$ SYM.
From this point of view type IIB strings on the pp-wave relate to type IIB 
strings on $AdS_5 \times S^5$ in a similar way to that which M(atrix) theory 
relates to M theory.
This enables us to explain many features of the BMN duality.
In particular, we feel it gives a better understanding of its exact nature.

The structure of the paper is as follows.
In section $2$ we discuss briefly the ``infinite momentum frame" (IMF) and 
``lightcone frame" (LCF). 
In section $3$ and $4$ we describe the Seiberg-Sen analysis of M(atrix) theory 
as DLCQ of M theory and of Matrix strings as DLCQ of type II string theories.
We stress features of the Seiberg-Sen limit that prove useful later.
In section $5$ we discuss the possibility of generalizing the DLCQ of string/M 
theory in the Seiberg-Sen approach to other spacetimes.
This leads us to the Penrose limit which is discussed in section $6$.
In section $7$ we draw the analogy between the Seiberg-Sen limit and the Penrose 
limit and put forward our suggested interpretation, namely, that they are 
essentially the same process.
In section $8$ we employ this point of view to type IIB string theory on $AdS_5 
\times S^5$. 
In section $9$ we use this DLCQ perspective to understand some of the features 
of the BMN duality such as the renormalization of the `t Hooft coupling.
In section $10$ we discuss in more detail some aspects of viewing BMN theory as 
DLCQ and point out an 
analogy between certain quiver theory operators discussed in \mrv\ and the 
matrices of M(atrix) theory.
In section $11$ we discuss the relation between the Seiberg-Sen and Penrose 
limits and the Maldacena limit used in AdS/CFT. 
We end with a summary in section $12$.

\newsec{Infinite momentum frame and lightcone frame.}

In this section we briefly review the issue of quantizing a system in the 
``Infinite momentum frame" (IMF) and in the ``Lightcone frame" (LCF) 
\refs{\imf,\susn} in flat space.

\subsec{IMF.}

Consider a system (e.g. of particles) in a Lorentz invariant theory in $d+1$ 
dimensions. Pick out one space dimension, say the $d^{th}$, call it the 
longitudinal direction and denote it by $x_{\parallel}$. The other space 
directions will be called ``transverse" and will be denoted by $x_{\perp}$. We 
thus use the basis $\{t,x_{\parallel},x_1,\dots,x_{d-1} 
\}=\{t,x_{\parallel},\vec{x}_{\perp} \}$.

The energy is given by the relativistic formula
\eqn\relen{E=\sum_{a}\sqrt{(p_a)_{\parallel}^2+(p_a)_{\perp}^2+m_a^2},} 
where the index $a$ runs over particles.
Boost now along the longitudinal direction in such a way that all particle 
states 
have a positive and ``Infinite" (i.e. larger than any other energy scale in the 
problem) longitudinal momentum
\eqn\ilm{|p_{\perp}|, m \ll |p_{\parallel}| \go \in} 
Such states will be called ``proper" states,  using the language of \susn.
Denote the total longitudinal 
momentum by $P_{\parallel}\equiv\sum_{a}(p_a)_{\parallel}$. 
Expanding \relen\ up to second order in $|{p_{\perp}\over p_{\parallel}}|$ and 
$|{m\over p_{\parallel}}|$ the energy formula becomes non-relativistic  
\eqn\nonren{E=|P_{\parallel}|+ \sum_{a} 
{(p_a)_{\perp}^2+m_a^2 \over 2|(p_a)_{\parallel}|},}
with $|(p_a)_{\parallel}|$ playing the role of non-relativistic mass.
All excitations not satisfying the condition above, i.e. whose momenta before 
the boost where comparable to $P_{\parallel}$, such as modes with negative 
$(p_a)_{\parallel}$, are called ``improper".
The improper modes decouple from the dynamics of the proper modes in the limit 
$P_{\parallel}\go\in$ since they 
become separated from them by an infinite energy gap. Compactifying the 
longitudinal direction on a circle of radius $R_s$ the longitudinal momentum 
is quantized $P_{\parallel}=N/R_s,$ and the IMF limit \ilm\ is 
$N \go \in.$ Thus, analyzing the system in the IMF simplifies the problem. 
\item{1.} The symmetry becomes Galilean. 
\item{2.} In the limit $P_{\parallel}=\in$ the improper modes decouple.

\subsec{LCF and DLCQ.}

Here one changes from $\{ t,x_1,\dots,x_d \}$ to ``lightcone" coordinates $\{ 
x^+,x^-,x_1,\dots,x_{d-1} \}$ via
\eqn\lco{x^{\pm}\equiv{t\pm x_d \over 2}.} Again denote 
$\{x_1,\dots,x_{d-1}\}\equiv \vec{x}_{\perp}$.
The conjugate momenta become
\eqn\lcm{p_{\pm}=i\partial_{\pm}=i(\partial_t \pm 
\partial_d)=E\mp P_d.} 
Choosing $x^+$ to play the role of time, the formula for the energy (the 
generator of $x^+$ translations) is (for the 
case $(p_a)_- \neq 0$)
\eqn\lce{H_{lc}\equiv P_+=\sum_a (p_a)_+ = \sum_{a} 
{(p_a)_{\perp}^2+m_a^2 \over 2(p_a)_-}.}
As in the IMF the longitudinal momentum $p_-$ plays the role of 
non-relativistic mass.
The symmetry group is not just the apparent $SO(d-1)$. It is the full 
Galilean group in $d$ space dimensions \susn. 
Also here it is convenient to compactify the null direction $x^-$ on a circle of 
radius $R_l$. The momentum modes are now quantized as $P_-={N \over R_l}.$ 
Quantizing a system in LCF with a null circle is called ``Discrete Light Cone 
Quantization" (DLCQ). Since $p_-$ is conserved (it is like the mass in Galilean 
mechanics) the Hilbert space decomposes into superselection sectors labeled by 
the 
integer $N$. Note that this analysis breaks down when $p_-=0$. The DLCQ does not 
obviously simplify the description of this ``zero mode" sector. 

Note that the LCF is simpler than the IMF in several ways.
\item{1.} \lce\ is exact for any value of $(p_a)_- \neq 0$ while \nonren\ was 
true 
only in 
the limit of infinite momentum. In the compact case this translates to the 
distinction between finite $N$ and infinite $N$.
\item{2.} Quantizing with respect to $x^+$ as the time and demanding to have 
a non negative Hamiltonian we see that for $(p_a)_- \neq 0$ we must have
\eqn\posmem{P_+=H_{lc}\geq 0\quad,\quad (p_a)_-> 0,} while in IMF improper modes 
(such as negative longitudinal momentum modes) decouple only in the limit $N \go 
\in$.

To summarize, all the advantages of the IMF appear now in each DLCQ sector with 
finite longitudinal momentum $N$. The two procedures are supposed to agree in 
the limit 
$N=\in$. In the next section we discuss the application of these 
old ideas \imf\ to string/M theory.

\newsec{M(atrix) Theory and The Seiberg-Sen limit}

The Seiberg-Sen (SS) analysis \refs{\seiberg,\sen} explains why the DLCQ of M 
theory with N units of 
longitudinal momentum is (as conjectured by BFSS \bfss\ and refined by Susskind 
\susn) 
the low energy limit of the 0+1 theory on the worldvolume of N $D0$ branes.
The analysis consists of two steps. The first step is the 
realization that the seemingly suspicious procedure of 
compactification along a null direction in flat space is best thought of as the 
limit of a spacelike compactification. This makes contact with the IMF. It is 
demonstrated that this limit leads to a theory of tensionless strings and is 
thus difficult to analyze. It is then observed that the relevant energies in the 
IMF also vanish in the limit. 
In fact, they vanish faster than the string tension. This leads to the second 
step that involves rescaling the Planck unit. This rescaling achieves a 
``focusing" on the relevant energy scale. It is then showed that with this 
rescaling one exactly ends up with the low energy limit of the 0+1 theory on the 
worldvolume of N $D0$ branes. 

We now review the Seiberg-Sen procedure in some 
detail.

\subsec{Step I - The null circle as a limit.}

We first discuss how the null circle is viewed as the limit of a spacelike 
circle.
The boost isometry of flat space with rapidity parameter $\alpha$ rescales the 
lightcone coordinates \lco\ as follows
\eqn\boost{x^{\pm}\go e^{\mp \alpha}x^{\pm}.}
Suppose $x_d$ is compact, i.e, $x_d \go x_d+2\pi R_s$ brings us back to the 
same point. In the boosted frame this translates into the following {\it 
combined} action which must be taken in order to return to the same point
\eqn\comb{x^+ \go x^++e^{- \alpha}\pi R_s \quad and \quad x^- 
\go x^-+e^{\alpha}\pi R_s.}
Sending  
\eqn\armin{\alpha\go\in\quad,\quad R_s \go 0\quad ,\quad e^{\alpha} 
R_s \equiv 2R_l \sim fixed,}  
the combined symmetry action reduces in the limit to $ x^- 
\go x^-+2\pi R_l$ so we get a null circle\foot{Note that at any 
finite boost parameter $x^-$ is {\it not} compact. Only in the 
exact limit does $e^{-\alpha}=0$ and we do not need to supplement the 
translation in $x^-$ with a translation in $x^+$.}.

It is easy to see why this limit gives a ``difficult" corner in the parameter 
space of string/M theory. Reducing M theory along $R_s$ we get a type IIA theory 
with the following coupling constant and string tension
\eqn\witten{g_A=({R_s \over l_p})^{3/2} \quad, \quad T=2\pi{R_s \over l_p^3},} 
so when $R_s \go 0$ we get tensionless strings.
The remedy also suggests itself, namely - a rescaling of $l_p$.

\subsec{Step II -  The rescaling of physical parameters.}

Denoting the components of the momentum vector $P^{\mu}$ in the non-boosted 
frame 
(where one has the spacelike circle) by a subscript $s$ for ``spacelike" and 
the (infinitely) boosted frame (the frame with the null circle) 
by a subscript $l$ for ``lightlike" we have at any intermediate finite 
boost parameter $\alpha$
\eqn\blue{\eqalign{E_s & ={1 \over 
2}e^{\alpha}\left(E_l+P_l+e^{-2\alpha}(E_l-P_l)\right) \cr
P_s & ={1 \over 2}e^{\alpha} \left( E_l+P_l-e^{-2\alpha}(E_l-P_l) \right) .}}
Thus, in the ``spacelike" frame all the momenta and energies are exponentially 
blue shifted with respect to the ``lightlike" frame. 
This is so because we are considering momentum modes along a vanishing circle.
However, the physics lies not in this diverging ``zero point" 
energy due to the boost but in the fluctuations. Equivalently, since we have a 
Galilean 
symmetry the mass is a constant parameter that does not take part in the 
dynamics.
Focusing on the fluctuations above the $D0$ brane mass we get using \lcm\lce 
\eqn\fluct{\Delta E_s=E_s-P_s=e^{-\alpha}(E_l-P_l)={R_s \over 2R_l} H_{lc},}
which vanish in the limit \armin. In order to focus on these as finite energy 
fluctuations we scale also the Planck 
mass\foot{Rescaled parameters will be denoted with tildes.} $\tilde{m}_p \go 
\in$ and the transverse size of the manifold $\tilde{R}_{\perp} \go 0$ along 
with the boost parameter $\alpha\go\in$ as
\eqn\ssca{\eqalign{\tilde{m}_p &\sim e^{\alpha \over 2}\go\in\cr 
\tilde{m}_p^2R_s &=m_p^2R_l \sim fixed \cr
\tilde{m}_p \tilde{R}_{\perp} &=m_pR_{\perp} \sim fixed.}}
The most easily extendible rationale behind this rescaling is that according to 
\fluct\ the relative energies scale like $R_s$ under further boosting. On 
dimensional grounds they should thus be proportional to $m_p^2R_s$ ($m_p$ being 
the only scale in M theory). 
Since the boost does not affect the transverse coordinates $R_{\perp}$ it makes 
sense to scale those like $\tilde{l}_p$ so that the transverse geometry in 
dimensionless units does not change during the limit.
This rescaling gives us the desired corner of parameter space, since now
\eqn\sswit{\eqalign{\tilde{g}_A&=(R_s\tilde{m}_p)^{3/2}=(R_s\tilde{m}_p^2)^{3/2}
\tilde{m}_p^{-3/2}\go 0 \cr 
\tilde{\alpha}'&=(R_s\tilde{m}_p^3)^{-1}=(R_s\tilde{m}_p^2)^{-1}\tilde{m}_p^{-1} 
\go 0.}}
Note also that if the transverse space contains a circle one can consider 
T-duality. Although the size of the transverse manifold vanishes like 
$\tilde{l}_p$ the size of the T-dual circle remains fixed in the limit. 
\eqn\tdual{{\tilde{\alpha}' \over \tilde{R}_{\perp}} = {1 \over 
(\tilde{R}_{\perp}\tilde{m}_p)(R_s\tilde{m}_p^2)} \sim fixed.}
This means that the natural physical description involves a T-duality. 
This is one way to see why the DLCQ of type IIA is given in terms of 
$D1$ strings and the DLCQ of IIB in terms of $D2$ branes.

\subsec{Scaling of dimensionful quantities.}

An essential part of the Seiberg-Sen analysis is the 
rescaling of the dimensionful quantity $m_p$. This point deserves an 
explanation. Why were we allowed to rescale the parameters of the theory? by 
doing so we naively change the theory. 
The point is that the scaling of dimensionful parameters is 
physically meaningless. The only physically invariant rescaling involve 
dimensionless parameters, so rescaling a dimensionful parameter is a way of 
focusing on some sector of the theory by sending a dimensionless parameter to 
zero. For instance we can use \nonren\ and \lcm\ to write the following relation 
for each particle in the IMF
\eqn\eqimdl{p_0-|p_{\parallel}|=p_+^{IMF}={p_{\perp}^2+m^2 \over 
2|p_{\parallel}|}={p_{\perp}^2+m^2 \over 2N}R_s.}
The Seiberg-Sen boost \boost\ takes $R_s \go 0$. So in order to focus on states 
of such energy one can introduce a mass scale $m_f$ and write
\eqn\eqimdlb{p_+^{IMF}=\left( {p_{\perp}^2+m^2 \over m_f^2}  \right) 
\left( {R_sm_f^2 \over 2N}\right) .}
Keeping the first parenthesis and $N$ fixed in the limit dictates that we must 
scale $m_f$ exactly as in \ssca. Using  \fluct\ we can now derive \lce.
So states with energy of order $m_f$ survive the 
rescaling as finite energy fluctuations. Those are the ``proper states".
States with energy of order $P_{\parallel}$ have infinite energy in $m_f$ units  
since ${P_{\parallel} \over m_f} \sim {1\over R_s m_f} \sim m_f \go \in$. Those 
are the ``improper modes" that decouple from the dynamics in DLCQ \susn. 
Thus, DLCQ achieves a simplification of the kinematics at the expense of 
focusing on the dynamics in a certain energy band. 

To summarize, the Seiberg-Sen limit in the context of M(atrix) theory consists 
of viewing the null circle as the end point 
of a limiting procedure during which the radius $R_s$ and the size of the 
transverse manifold vanish\foot{It is worth mentioning here that 
since the DLCQ of M theory involves a compactification on a vanishing 
circle supergravity is no longer a good description. This explains \Helling\ why 
various comparisons made in the literature between perturbative results in 11 
dimensional DLCQ supergravity and 
calculations in the Matrix model turned out to agree only for quantities 
protected by supersymmetry. (for a recent review see \wati).} in such a way so 
as to keep
\eqn\sscale{\eqalign{R_l & \equiv ({\tilde{m}_p \over m_p})^2 R_s \sim fixed \cr 
R_{\perp} & \equiv  {\tilde{m}_p \over m_p} \tilde{R}_{\perp} \sim fixed.}}

\newsec{Matrix strings and the Seiberg-Sen limit.}

The original Seiberg-Sen analysis was done for the DLCQ of M theory. The DLCQ 
of string theory was treated later by \refs{\dvv,\bs,\motl}. The 
essential new feature of this analysis is the $9-11$ flip. In this section we 
briefly review Matrix strings.

\subsec{Matrix strings.}

In order to describe the DLCQ of type IIA string theory with parameters $g_A, 
\alpha '$ having longitudinal momentum $P_-={N \over R_l}$ along a null circle 
$x^-\sim 
x^-+2\pi R_l$ we need to perform a lift to 
M theory along {\it another} circle $R_9=g_Al_s$ and get M theory with 
$l_p=g_A^{1 \over 3}l_s$. 
Reducing now along the {\it null} circle we get a {\it different} IIA theory 
with 
(hatted parameters) 
\eqn\mpara{R_l=\hat{g}_A\hat{l}_s, \qquad, \qquad 
l_p=g_A^{1/3}l_s=\hat{g}_A^{1/3}\hat{l}_s}
along with $N$ $D0$ branes.
The end result (after one T-duality) is the low energy theory on N D1s wrapping 
a circle with parameters
\eqn\dualr{R_{D1}={\alpha ' \over R_l}\quad,\quad g_{YM}={R_l \over g_A \alpha 
'}.}  An analogous treatment is done for 
the DLCQ of type IIB on a circle by first T-dualizing to type IIA. 

\subsec{Seiberg-Sen analysis of Matrix strings.}

Now we turn to the Seiberg-Sen point of view. The null circle with radius $R_l$ 
is viewed 
as the limit of a vanishing spacelike circle of radii $R_s \go 0$ with 
$2R_l=e^{\alpha}R_s \sim fixed$. 
Repeating the above analysis but now the compactification 
done on the spacelike circle $R_s$ we get essentially the same answer but with 
$R_l$ replaced by $R_s$, namely
\eqn\dualrb{R_{D1}={\alpha ' \over R_s}\quad ,\quad g_{YM}={R_s \over g_A \alpha 
'}. }
Now the limit seems to give a singular result. The D0 branes are transverse to 
a vanishing circle and the YM coupling vanishes as well. 
The solution is to rescale $l_p$ as in \sscale, i.e. 
\eqn\msres{\tilde{l}_p \sim \sqrt{R_s} \quad or \quad R_s \tilde{m}_p^2 \sim 
fixed.} 

The $9-11$ flip gives us an interesting way of rescaling 
the 11 dimensional Planck length that has a clear physical 
interpretation in the original IIA string theory which we are describing in 
DLCQ.
Looking at \mpara\ we can achieve the appropriate rescaling of $\tilde{l}_p$ by 
rescaling the string tension in the original IIA in following 
manner
\eqn\resc{l_s\go \tilde{l}_s \sim \sqrt{R_s} \qquad, \qquad g_s \sim 
fixed.}
In this scaling both quantities in \dualrb\ are being held fixed 
in the limit.
If we have also transverse directions, keeping the geometry fixed in {\it 
string} units one needs to define $\tilde{R}_{\perp} \sim \tilde{l}_s$ such that 
$R_{\perp} / l_s=\tilde{R}_{\perp}/ \tilde{l}_s$. 
These scaling in string theory exactly implement \sscale.
Note that $R_9=g_Al_s$ which after the $9-11$ flip is just a transverse 
circle indeed scales like $l_s$ as it should.
The same analysis with the rescaling \resc\ works also for the DLCQ of type IIB 
on a circle.

To summarize, in both the DLCQ of M theory and of type IIA/B string theory the 
Seiberg-Sen limit is being effectively implemented by correctly scaling the 
fundamental 
length scale of the theory which we will denote here by $l_f=m_f^{-1}$. In all 
cases the scaling is a decoupling limit 
\eqn\mfscale{\tilde{m}_f \go \in,} along with the 
vanishing limit of the spacelike circle $R_s \go 0$ while keeping
\eqn\genscale{\eqalign{R_l & \equiv ({\tilde{m}_f \over m_f})^2R_s \sim fixed 
\cr R_{\perp} & \equiv  \ {\tilde{m}_f \over m_f} \ \tilde{R}_{\perp} \ \sim 
fixed.}} 

In the next section we suggest how one can apply this procedure to more general 
string backgrounds.

\newsec{An attempt to generalize DLCQ.}

The analysis presented in the last section is not directly applicable to any 
background and tacitly assumes 
that the space is the product of flat space (where time and the null circle 
reside) and a general compact manifold of some generic size $R_{\perp}$
\eqn\mflat{R^{1,p}\times {\it M}_{\perp}^{9-p}.}
In other words it assumes there is a boost isometry.
In this section we will try to suggest a possible generalization of DLCQ to more
general spacetimes. 

In order to generalize the concept of DLCQ to backgrounds not of the form 
\mflat\ we restate the basic idea behind DLCQ. We are describing a theory 
symmetric under a boost in the $\{ t,x_d \}$ plane. Thus, we can use this 
symmetry to describe the physics from the LCF. 
In other words, since different inertial observers are related by the boost 
symmetry 
we are free to choose the observer that gives the simplest description.
So what happens in more general spacetimes? In a theory of gravity 
(such as string/M theory) {\it all} observers 
are equally fit to describe the physics. Typically the simplest description 
appears in the frame of the appropriate generalization of ``inertial observer" 
to curved spaces, namely, in the frame of a freely falling observer.
Those observers move along geodesics. 
What made the LCF/IMF description simple was the introduction of a large 
quantity, namely, the longitudinal momentum which boosted the observer 
asymptotically close to the speed of light. 
In a general spacetime we should adopt a local version of this statement, 
namely, ``moving asymptotically close to a {\it null} geodesic". The same 
physical reasoning suggests that a simple 
theory might emerge if one also rescales the fundamental scale of the theory 
appropriately.

Our suggestion immediately raises several questions. 
\item{1.} Does such a limit exist?
\item{2.} Why is this lightcone quantization discrete? where is the circle in a 
spacetime that does not have a null isometry like flat space?
\item{3.} Does such a procedure reduce to the usual DLCQ in flat space? for 
instance can we get M(atrix) theory that way?

The central point of this paper is to argue that indeed this is a well defined 
procedure known as ``the Penrose limit of a spacetime". In the next 
section we describe the Penrose limit and show that it generalizes in a precise 
way the Seiberg-Sen prescription. Later on we will address the other questions 
presented above.

\newsec{The Penrose limit.}

The Penrose limit \refs{\penrose,\guven} is defined by focusing on the geometry 
near a null geodesic.
Locally, in the neighborhood\foot{The neighborhood must not contain conjugate 
points.} of a null geodesic one can introduce the 
following set of coordinates $Y\equiv\{ y^{\pm},y^i \}$ such that the line 
element is given by
\eqn\penco{ds^2=g_{\mu\nu}dy^{\mu}dy^{\nu}=-2dy^-[dy^++A(Y)dy^-+B_i(Y)dy^i]+C_{i
j}(Y)dy^idy^j,}
where $i,j=1,\dots,8$ and $C_{ij}$ is a positive definite symmetric matrix.
The close-by null geodesics are parameterized by $y^-,y^i=const$ and the affine 
parameter along them is $y^+$. The ``original" null geodesic on which we focus 
is at $y^-=0$.

The next step in taking the Penrose limit is to blow up this neighborhood to 
become the whole space\foot{This is reminiscent of the ``near horizon limit" of 
Maldacena. We discuss this point later on.}.
This is done by introducing an auxiliary dimensionless parameter 
$\Omega\ $ and defining a rescaled set of coordinates $X\equiv\{ x^{\pm},x^i \}$ 
\eqn\penscale{\eqalign{x^+&\equiv y^+,\cr x^-&\equiv \Omega^2 y^-,\cr x^i& 
\equiv \Omega y^i.}}
Penrose then tells us that the rescaled metric 
\eqn\resmet{G_{\mu\nu}(X)\equiv \Omega^2 g_{\mu\nu}(Y)}
has the following well defined limit\foot{This is easily seen by expanding 
$C_{ij}(Y)dy^idy^j=C_{ij}(x^+,{x^- \over \Omega^2},{x^i \over \Omega}){dx^i 
\over \Omega}{dx^j \over \Omega}$ in powers of ${1 \over \Omega}$.} as $\Omega 
\go \in$ 
\eqn\penlim{ds^2=G_{\mu\nu}dx^{\mu}dx^{\nu}=-2dx^+dx^-+C_{ij}(x^+)dx^idx^j,}
where $C_{ij}(x^+)\equiv C_{ij}(x^+,0,\vec{0})$.
This is the pp-wave in Rosen coordinates. 
One can change to Brinkman coordinates in which the line element is given in the 
more familiar form 
\eqn\genppw{ds^2=-4dz^+dz^--H_{ij}(z^+)z^iz^j(dz^+)^2+dz^idz^i.}
We do not concern ourselves with other background fields such as 
gauge fields or p-form fields. All those can be scaled appropriately in the 
Penrose limit \guven.
In Rosen coordinates it is obvious that the pp-wave has many isometries. Note 
that there is 
always the null isometry along the coordinate in the ``minus" direction. 
We next argue how one can naturally take the Penrose limit in such a way that 
effectively 
compactifies the null isometric direction so that in the end we get a null 
circle.

\subsec{Penrose limit + identifications, or from null isometry to a null 
circle.}

Near the null geodesic there is a coordinate system of the form \penco.
Lets us introduce a local time and space \co\ 
\eqn\stco{t=y^++y^-\quad,\quad s=y^+-y^-.}
Assume that the metric has an isometric direction\foot{Note that this restricts 
the 
discussion to spacetimes that have locally at least one spacelike and one 
timelike killing vectors. Spacetimes that do not have those minimal requirements 
would probably be too difficult to analyze in any case.}$^,$\foot{If this is not 
the case, 
perhaps one can still approximate the space by another one that does have this 
isometry, in such a way that the difference between the metrics on the two 
spaces vanishes in the limit. We do not pursue such a generalization here.} with 
some component 
along $s$. Compactifying along that isometric direction involves 
\eqn\xcom{s \go s+ c_{\ \parallel},} with $c_{\ \parallel}$ some constant. Of 
course, $s$ need not be the isometric direction itself so the actually symmetry 
transformation may also involves an action on some other ``transverse" spacelike 
coordinates $y^i$ which we will symbolically denote here as
\eqn\symbtr{\alpha_{\perp}\go\alpha_{\perp} +c_{\perp},} with $c_{\perp}$ some 
other appropriate constants. The radius of the circle is
\eqn\radci{R_s \sim  c_{\ \parallel}\sim c_{\perp}.}
\xcom\ induces the following combined action on the lightcone \co\
\eqn\combb{y^+\go y^++{1 \over 2} c_{\ \parallel}\quad,\quad y^-\go y^--{1 \over 
2} c_{\ \parallel}.}
In the Penrose limit
the transverse \co\ get rescaled \penscale\ so let us define 
$\beta_{\perp}\equiv\Omega\alpha_{\perp}$ to be the rescaled spacelike \co\ 
involved in the circle identification. This gives
\eqn\combbc{\eqalign{x^+&\go x^++{1 \over 2} c_{\ \parallel}, \cr x^-&\go x^--{1 
\over 2} c_{\ \parallel} \Omega^2 \cr
\beta_{\perp}&\go\beta_{\perp}+c_{\perp}\Omega.}}
The Penrose limit sends $\Omega \go \in$. If we {\it supplement} the Penrose 
limit by the following {\it rescaled identification}, namely, \xcom\ together 
with 
\eqn\rides{c_{\ \parallel} \go 0 \quad,\quad c_{\ \parallel} \Omega^2\sim 
fixed,}
we get in the limit that the combined action degenerated to an action only on 
the $x^-$ direction
\eqn\nulcic{x^-\sim x^-+2\pi A_l,} where we have defined ${1 \over 2}c_{\ 
\parallel} \Omega^2\equiv 2 \pi A_l.$ Namely, we get a null circle. 
The following general argument was demonstrated in 
specific cases when discussing the Penrose limit of orbifolds of 
the type $AdS_5 \times S^5/Z_M$ \refs{\mrv,\bdhio}. These models will 
be addresses in detail later on and will make the above abstract procedure 
clearer.

\newsec{Penrose limit, DLCQ and the Seiberg-Sen limit.}

We saw that any spacetime has a Penrose limit. This limit is 
achieved by focusing on the neighborhood of a null geodesic and rescaling 
the coordinates in a way that blows up the neighborhood to become the whole 
space. 
The rescaling is universal \penscale
\item{A.} The time coordinate does not get rescaled. 
\item{B.} The null coordinate gets rescaled {\it quadratically} in an auxiliary 
parameter.
\item{C.} The spacelike transverse coordinates get rescaled {\it linearly} in 
that parameter.

We further argued how one can use a slight generalization of the Penrose limit 
so as to naturally end up with a null circle.
This generalization involves a limit of discrete identifications that produce a 
vanishingly 
small radius \rides\ in the non-rescaled \co\ $Y$. The circle vanishes with the 
same rescaling as the null coordinate. 
These rescalings are easily recognized to be identical to those done in the 
Seiberg-Sen 
treatment of the DLCQ of string/M theory, namely \sscale\ with the 
identification 
\eqn\plom{\Omega\equiv{\tilde{m}_p \over m_p}.}
 
We are thus led to the following statement.
The analogue of M(atrix) theory for string/M theory on general curved 
backgrounds 
is given by string/M theory on the Penrose limit of that background.
The Penrose limiting procedure supplemented by appropriately rescaled discrete 
identifications gives the DLCQ. 
The ``usual" Penrose limit is the decompactification limit, or DLCQ in the 
infinite longitudinal momentum limit. 
The Penrose limit automatically achieves both steps in the Seiberg-Sen 
procedure, namely, 
describing the system from the point of view of an observer moving 
asymptotically close to the speed of light and 
rescaling parameters so as to focus on finite energy fluctuations.
We believe that this ``focusing" property of DLCQ is being mirrored in the 
geometry by the fact that the Penrose limit blows up the neighborhood of a null 
geodesic to become the whole space and ``throws the rest" to infinity\foot{This 
is reminiscent of what happens to the region outside the ``throat" in 
Maldacena's limit.}.
This corresponds to the decoupling of heavy modes in DLCQ according to the usual 
UV/IR relation in AdS/CFT \witsus.

We feel this answers the first two questions posed above. It is also 
straightforward to see that this procedure reduces to the Seiberg-Sen limiting 
procedure if one considers the Penrose limit of 11 dimensional flat space 
compactified on a circle.

\newsec{DLCQ of strings in $AdS$ space.}

In this section we finally reach the model that motivated this line of research.
We argue that type IIB string theory on a Penrose limit of the orbifold 
background $AdS_5 \times S^5/Z_M$ together with an appropriate scaling of the 
rank of the orbifold group $M$ (see e.g. \refs{\mrv,\bdhio}) is a {\it DLCQ of 
type IIB on} $AdS_5 \times S^5$.
The lightcone quantization here is discrete since the resulting pp-wave space 
has a null circle. This theory is analogous to Susskind's finite $N$ version of 
M(atrix) theory \susn.
The limit of infinite longitudinal momentum corresponds to the Penrose limit of 
$AdS_5 \times S^5$, namely, to the BMN theory. This theory is analogous to the 
BFSS M(atrix) theory \bfss.

\subsec{The DLCQ of $AdS_5 \times S^5$.}

Just as in flat space, it is best to start from the DLCQ, namely with finite
longitudinal momentum. To that end we need to discuss strings on the Penrose 
limit of
$AdS_5 \times S^5/Z_M$. Those have been studied in e.g. \refs{\mrv,\bdhio}.
We choose to focus on the case where the null geodesic does not pass through 
singular points of the orbifold. This Penrose limit results in the maximally 
supersymmetric type IIB pp-wave.
Following \mrv\ we write the metric on $AdS_5 \times S^5/Z_M$ as
\eqn\orbmet{\eqalign{ds^2=&R^2 [ -\cosh^2\rho \ dt^2 \ + d \rho^2 \ + \ \sinh^2 
\rho \ d \Omega_3^2 \ +\cr \  &d\alpha^2 \ + \ \sin^2 \alpha\ d \theta^2\ + 
\cos^2 \alpha \left( d \gamma^2 \ + \cos^2 \gamma\ d \chi^2\ + \sin^2 \gamma\ d 
\phi^2 \right) ] ,}}
where the first line is the $AdS_5$ metric in global coordinates. The second 
line is the metric on $S^5$ embedded in $R^6 \simeq C^3$ with coordinates
\eqn\compco{z_1=R \sin\alpha\, e^{i\theta}, \quad z_2=R\cos\alpha \cos\gamma\,
e^{i\chi}, \quad z_3=R\cos\alpha \sin\gamma\, e^{i\phi}. }
The orbifold action identifies any point with the point resulting from the 
combined action 
\eqn\orbif{\chi \go \chi +{2 \pi \over M} \qquad, \qquad \phi 
\go \phi -{2 \pi \over M}.}
This choice of the metric and of the orbifold action explicitly break the 
$SO(6)$ isometry group of the 5 sphere into $U(1) \times SO(4)$, where the 
$U(1)$ is parametrized by $\theta$.

In order to take the Penrose limit we choose to focus on the following null 
geodesic 
\eqn\nulbmn{\chi=t \quad,\quad \rho=\alpha=\gamma=0,} and rescale the \co\ in 
its neighborhood as follows
\eqn\resco{\eqalign{ x^+  &\equiv {1 \over \mu}{t+\chi \over 2} \cr x^-  
&\equiv \mu R^2  {t-\chi \over 2} \cr
r \equiv R\rho\quad ,\quad \omega &\equiv R\alpha\quad , 
\quad y \equiv R\gamma\ ,}} with $\mu$ an arbitrary positive parameter of mass 
dimensions.
Making the substitution \resco\ and taking the limit $R\go \in$ one 
is left with the maximal supersymmetric pp-wave background 
\eqn\mspp{ds^2=-4dx^+dx^--\mu^2 z^2(dx^+)^2+dz^2,} where 
\eqn\zisq{dz^2\equiv 
\sum_{i=1}^8dz^idz^i=dr^2+r^2d\Omega_3^2+d\omega^2+\omega^2d\theta^2+dy^2+y^2d
\phi^2} denote 
the 8 flat transverse coordinates (four originating from the $S^5$ and four from 
the $AdS_5$ factors).
Again, we suppress the RR-form since it is of no importance to our discussion.
However, as opposed to \bmn\ here $x^-$ can be made compact by appropriately 
scaling $M$.
This is so since the identification \orbif\ sends any point to an identical 
point by the combined action
\eqn\orcom{x^+\go x^++{\pi \over \mu M}\quad, \quad x^-\go 
x^-+{\mu R^2\pi \over M}\quad, \quad \phi\go\phi-{2 \pi \over M}.}
If we scale together 
\eqn\mscale{R,M \go \in \quad, \quad {\mu R^2 \over M}\equiv 2R_l \sim 
fixed,} we see that {\it in the limit} any point is mapped to an identical point 
by only 
sending $x^-\go x^-+2\pi R_l,$ i.e. $x^-$ parameterizes a null circle of 
radius $R_l$.

However $R$ is a dimensionful 
parameter and thus there is no physics in the claim that $R\go \in$. The only 
physically meaningful scaling involve dimensionless quantities. Since we are 
talking about string theory in $AdS_5\times S^5/Z_M$ we can equivalently choose 
$R/l_s$ or $R/l_p$ as the dimensionless parameter, since the string coupling is 
constant. 
Choosing $\Omega\equiv {R \over l_s}\sim (g_sN)^{1 \over 4}$ the Penrose limit 
can 
equally be taken by 
\eqn\eqscb{ N\go\in \quad,\quad g_s \sim fixed,} which sends
\eqn\penlima{\alpha'\sim {1 \over 
\sqrt{N}}\go 0\quad,\quad g_s\sim fixed.}

So let us identify 
\eqn\ident{{1 \over M} \equiv \mu R_s \quad, \quad \Omega \equiv {R 
\over l_s}\sim m_s.}
The relation \mscale\ is equivalent to saying 
\eqn\resei{R_s\go 0 \quad , \quad m_s \go \in \quad, \quad 
R_sm_s^2\sim fixed\quad, \quad g_s\sim fixed.}
This is the same limit as in the Matrix string \resc.
The transverse neighborhood, whose size (generally denoted by $R_{\perp}$) in 
this case is given by the three radii $\rho,\alpha,\gamma$ also scale 
appropriately since $R_{\perp}m_s\sim fixed$ in the limit. So we recover 
\genscale.

Now let us look back at the gauge theory side. 
Here ${1 \over M}$ is the periodicity of the difference angle ${\chi-\phi\over 
2}$ inside the $S^5/Z_M$.
The lightcone Hamiltonian and longitudinal momentum are 
\eqn\bmndic{\eqalign{H_{lc}=p_+&=\mu(\Delta -J) \cr p_-&={ \Delta +J \over \mu 
R^2}.}}
Notice that the orbifold identification \orbif\ is not only within the great 
circle parameterized by $\chi$ along which we boost by \resco.
So let us divide the current as follows \mrv
\eqn\jay{J=-i\p \chi=J^++J^-,} where we define
\eqn\dejey{J^{\pm}\equiv -{i \over 2}(\p \chi \pm \p \phi)\equiv -i{\p \over \p 
\varphi^{\pm}},} and
\eqn\plminph{\varphi^{\pm}=\chi \pm \phi.}
This choice of \co\ is the geometric manifestation of the group theory statement 
$SO(4)=SU(2)_+\times SU(2)_-$ \bdhio, where the currents \dejey\ generate the 
Cartan subalgebra. In particular the eigenvalues of $J^{\pm}$ are half integral.

The orbifold identification \orbif\ acts only on $\varphi^-$ by
\eqn\orvar{\varphi^- \sim \varphi^-+{\pi \over M}.} Its effect is to project on 
the subspace of states periodic over $1/2M$ of the full period, namely, only 
states with 
\eqn\keye{J^-=M\times(2k),} with $k$ half integral, survive the orbifold 
projection.

To get finite quantities we send following \bmn\ $\Delta \sim J\sim\sqrt{N}\go 
\in $.
But we saw \mscale\ that if we want a null circle also $M\sim \sqrt{N}$ so $J/M 
\sim fixed$. Therefore, we can keep finite quantum numbers for $k$ and $J^+$ 
and still get the Penrose limit since $M \go \in$.
Let us define the following integer number 
\eqn\ququ{q\equiv 2k.}
From \bmndic\ we get using \keye\ and \mscale\ that in the Penrose limit
\eqn\nulmpp{p_-\sim {2J \over \mu R^2}={2(qM+J^+)\over \mu R^2}\go{q \over 
R_l}.}
So $q$ is the quantum number denoting the longitudinal momentum along the null 
circle in the resulting pp-wave background. Remember that in M(atrix) theory, 
the number $N$ of $D0$ branes was the number of longitudinal momentum quanta. We 
thus suggest to interpret $q$ as the analogue of the number of $D0$ branes in 
Susskind's finite $N$ reformulation of M(atrix) theory \susn.
The diverging quantum number $J$ corresponds to the diverging mass of the $D0$ 
brane. 

In the Penrose limit of $AdS_5 \times S^5$ studied by BMN no orbifold was taken 
so this is the special case $M=1$. 
This means that $q\sim J\go \in$. In other words the 
original BMN paper analyzed the DLCQ of string theory on $AdS_5 \times S^5$ in 
the limit of infinite longitudinal momentum. This is analogous to the BFSS  
M(atrix) theory. 
To summarize, the study initiated in \bmn\ of strings on the Penrose limit of 
$AdS$ space is a concrete realization of the generalized DLCQ procedure proposed 
here.

\newsec{Known facts about BMN from the DLCQ perspective.}

In this section we employ the DLCQ perspective to understand some features of 
the BMN duality.

\subsec{``Renormalization" of coupling constants.}

The effective expansion parameter in the sector dual to type IIB on the pp-wave 
was shown\refs{\bseven,\plefka} to be 
not the `t Hooft coupling $\lambda=4\pi g_BN$, which diverges in the Penrose 
limit, but rather 
\eqn\lampr{\lambda'\equiv {\lambda \over  J^2} \sim fixed.}
Also the effective genus expansion parameter for the Feynman graphs is 
``renormalized" 
\eqn\genex{g_2={J^2 \over N}.}
This phenomenon got a convincing combinatorial explanation by studying the 
relevant Feynman graphs (see e.g. \bseven,\plefka).
We now show that this renormalization is a direct manifestation of the 
Seiberg-Sen rescaling.
Since we are really describing the DLCQ of strings in 
$AdS_5 \times S^5$ we can use the standard AdS/CFT relation \juan
\eqn\adsc{({R \over l_s})^4=\lambda=4\pi g_B N.}
In AdS/CFT we keep this quantity fixed (`t Hooft 
limit) but here we are sending this 
dimensionless quantity to infinity. BMN tell us to send $R \go \in$ 
and scale $J \sim R^2$. Physically this is the same as setting $R=1$ and 
scaling $\alpha' \sim {1 \over J} \go 0$ while keeping $g_B$ fixed. 
It follows that 
\eqn\adscb{g_B\sim {1 \over N \alpha'^2} \sim {J^2 \over N}=g_2 \sim fixed.}
We see that \genex\ is the original IIB string coupling on $AdS_5 \times S^5$.
From this relation \lampr\ follows accordingly.

\subsec{Non-planar diagrams at infinite $N$ and second quantization.}

A closely related fact is that even at the large $N$ limit the genus expansion 
is nontrivial, and one needs to take into account diagrams of all genera. 
In fact, the string 
coupling \adscb\ is non vanishing (it is the original IIB string coupling of the 
$AdS_5 
\times S^5$ background). 
This phenomenon seems reminiscent of a central property of the DLCQ of M theory, 
namely, that M(atrix) theory is argued to be a second quantized theory. It is 
tempting to assume that also the DLCQ of string in $AdS$ via the Penrose limit 
gives rise to a second quantized theory. Thus one should expects that even in 
the strict $N = \in$ limit strings will interact, and the effect of higher 
genera will be indispensable.

\subsec{Pp-wave algebra, Inonu-Wigner contraction and the Galilei group.}

One of the simplifying features of DLCQ and IMF in flat space is the appearance 
of a Galilean symmetry instead of the Lorentz symmetry.
An analogous statement exist also for the DLCQ of strings on $AdS_5 \times S^5$. 
The symmetry algebra of the Penrose limit of $AdS_5 \times S^5$ is an 
Inonu-Wigner (IW) contraction \iw\ of the symmetry algebra of the full space, 
namely, $SO(2,4) \times SO(6)$\foot{We discuss only the bosonic part of the 
symmetry algebra. This can be generalized to the supersymmetric 
case \contjap.}. 
The Inonu-Wigner contraction is also the procedure
by which one gets the Galilei group as a limit of small velocities (or infinite 
rest mass) from of the Lorentz group \weinberg. 
This is exactly what happens in the Penrose limit. By moving 
close to a null geodesic we give a diverging lightcone ``rest mass" to all the 
particles \lce.
Thus, the IW contraction of the symmetry algebra in the Penrose limit should be 
understood as the exact analogue of the appearance of the Galilean symmetry in 
flat space\foot{Note that this gives a physically intuitive ``explanation" of 
the 
fact \genscale\ that the transverse \co\ scale linearly (momentum is linear in 
the velocity for small velocities) and the longitudinal quadratically (the 
energy is quadratic in small velocities).}$^,$\foot{A related point noticed in 
\mrv\ 
is that performing a T-duality in the quiver theory space leads to a 
non-relativistic string theory (NRST)\nrst.}.

\subsec{Decoupling of negative modes and the BPS condition.}

Due to the BPS condition of the $N=4$ supersymmetry algebra, $\Delta \geq |J|,$ 
both $p^{\pm}$ in \bmndic\ are positive.
It is clear that the lightcone Hamiltonian should be positive, but why is the 
longitudinal momentum positive? This is one of the features of DLCQ (see 
\posmem). This is analogous to the reason that in M(atrix) theory one has only 
$D0$ and no anti-$D0$ branes.
In fact, BMN argued that the insertion of a $\bar{Z}$ impurity decouples in the 
Penrose limit. This is the statement that in IMF and DLCQ the modes that have 
negative longitudinal momentum (in this case negative $J$) decouple.

\subsec{The parameter $\mu$.}

This (nonphysical) parameter appears due to a rescaling symmetry of the pp-wave.
In the language of DLCQ this parameter reflects the fact that we have an 
infinite quantity in the problem, namely the infinite momentum. We are free to  
``rescale" infinity by a positive number. The extra freedom of choosing $\mu$ is 
analogous to the freedom of performing additional boosts in the IMF.

\subsec{Supersymmetry.}

The BMN pp-wave background has 32 supercharges, however, only 16 of them are 
linearly realized \ppw. This is reminiscent of the fact that M(atrix) theory 
gives a DLCQ description of a theory with 32 supercharges, in terms of $D0$ 
branes that are ${1 \over 2}$ BPS.

\newsec{A closer look at BMN as DLCQ.}

In this section we take a closer look at the BMN theory as a DLCQ of type IIB 
string theory on $AdS_5 \times S^5$. We also point 
out an analogy between the matrices of M(atrix) theory and a certain type of 
operators in the $N=2$ quiver gauge theory dual to type IIB strings on $AdS_5 
\times S^5/Z_M$.

\subsec{Excited string states.}

Using AdS/CFT \witads\ one can predict the following general 
behavior of conformal dimensions of single trace operators in $N=4$ SYM in the 
supergravity approximation
\eqn\sugrap{1\quad \ll \quad\lambda=g_BN={R^4 \over \alpha'^2}\quad \ll \quad N= 
{R^4 \over \l_p^4}.}
\item{A.} dimensions of order $\sim 1$ correspond to KK modes of the reduction 
on the $S^5$. They are all BPS states in $10$ dimensions, and thus all 
correspond to
chiral primaries in the CFT. Those states exist up to $\Delta=N$.
\item{B.} dimensions of order $m_s \sim \lambda^{1 \over 4}$ corresponds to 
excited string states.
\item{C.} dimensions of order $1/g_B\sim N$ correspond to D-branes.
\item{D.} dimensions of order $1/g^2_B\sim N^2$ correspond to NS5-branes. 

Let us look at the energy formula in BMN
\eqn\enbmn{H_{lc}=\mu\sum_{n=-\in}^{\in} N_n \sqrt{1+{n^2 \over (\mu P_- 
\alpha')^2}}.}
In the limit the contribution of each oscillator to the anomalous dimension is 
given by
\eqn\contos{(\Delta-J)_n=\sqrt{1+{4 \pi g_BNn^2 \over J^2}}.}
The $n=0$ sector describes the supergravity modes. We will discuss those 
momentarily.
The excited string states correspond to $n\neq 0$.
Note that in the region where the second term in \enbmn\ dominates one gets back 
the usual string spectrum in flat space where we have the relation 
\eqn\strfl{E^2 \sim Tn,} with $T$ the string tension.
Looking at the same limit in \contos\ we identify 
\eqn\tilam{T \sim \sqrt{\lambda'},} where \lampr\ $\lambda'=\lambda/J^2$.
Using the usual translation formula between energies and dimensions  
\eqn\deltar{\Delta \sim RE,}
and the rescaling used by BMN $J\sim R^2$ we get 
\eqn\endel{\Delta \sim RE \sim R\sqrt{T} \sim R(\lambda')^{1 \over 4}\sim 
\sqrt{J}(\lambda')^{1 \over 4}=(\lambda)^{1 \over 4}.}
This is the correct region where one expects strings to appear in $AdS_5 \times 
S^5$. We interpret this as evidence for the relation between the string 
excitations in the pp-wave and the string excitations in $AdS_5 \times S^5$.

\subsec{Supergravity modes.}

The $n=0$ sector was identified in \bmn\ as corresponding to the supergravity 
modes propagating in the plane wave geometry. From the point of view described 
here it should contain information also about supergravity modes in $AdS_5 
\times S^5$. 
At first sight this seems impossible for several reasons. The BMN states consist 
only of the part of the spectrum of the $N=4$ theory that has divergingly large 
dimensions. So how can we even expect to see the known states with small 
dimensions of order $\sim 1$. Also, the representations of the pp-wave algebra 
are very different from the representations of $SO(2,4)\times SO(6)$.
We believe the answer lies in the way DLCQ ``blows up" bands in the 
spectrum. This operation ``distorts" the spectrum but does not change the 
Hilbert space. Indeed, the action of the Seiberg-Sen/Penrose limit 
on the Hilbert space can be understood in two equivalent ways (the usual 
passive/active descriptions) 
\item{A.} Keeping the $AdS_5 \times S^5$ symmetry algebra, namely we are still 
classifying states according to scaling dimensions and Lorentz quantum numbers, 
but we are looking at states with diverging dimensions and $R$ charge. This is 
the point of view presented in BMN.    
\item{B.} We keep the full Hilbert space of supergravity modes on $AdS_5 \times 
S^5$ 
but we contract the $AdS_5 \times S^5$ symmetry algebra to the pp-wave algebra.
The Hilbert space is always there, but the symmetry operators get rescaled so 
that many states end up with either zero or $\in$ eigenvalues.

Let us illustrate this point using an example. Consider the simplest 
group contraction, namely, that of $SU(2)$
\eqn\sutwo{[J^3,J^{\pm}]=\pm J^{\pm} \quad,\quad [J^+,J^-]=2J^3,}
with Casimir
\eqn\casim{J^2=J^+J^-+(J^3)^2-J^3.}
The group contraction amounts to the rescaling  
\eqn\suom{J^3=\Omega \tilde{J}^3\quad,\quad \Omega \go \in.}
In the limit $\Omega\go\in$ $\tilde{J}^3$ becomes a central element since
\eqn\centrael{[\tilde{J}^3,J^{\pm}]=\pm {J^{\pm} \over \Omega}\go 0.}
So the Hilbert space decomposes into sectors labeled by the $\tilde{J}^3$ 
quantum number. Let us denote this quantum number by $\tilde{m}$. The parent 
$J^3$ quantum number will be denoted by $m$ and the relation is
\eqn\mmm{\tilde{m}={m \over \Omega}.}

Now focus on a sector labeled by a given $\tilde{m}$.
In this sector the $J^{\pm}$ satisfy the following relation
\eqn\jeypm{[J^+,J^-]=2\tilde{m}\Omega .}
We recognize this to be the algebra of a harmonic oscillator by defining
\eqn\tiljpm{J^{\pm}=\sqrt{2\tilde{m}\Omega}\tilde{J^{\pm}},} which satisfy in 
the limit $\Omega \go \in$
\eqn\sutwocon{[\tilde{J}^3,\tilde{J}^{\pm}]=0 
\quad,\quad[\tilde{J}^+,\tilde{J}^-]=1.}
So now we have the algebra of a harmonic oscillator and another central element.
Clearly the representations of the harmonic oscillator are very different from 
representations of $SU(2)$. For one thing, they are infinite dimensional. But 
what really happened here is just a relabeling of the Hilbert space. 
Let us look at the Casimir \casim\ in the tilded variables. 
In the limit we are considering the $\tilde{m}$ is just a number in each 
superselection sector, so we can just as well label the representations by 
\eqn\enen{\widehat{N}=\tilde{J}^+\tilde{J}^-,} which is the number operator, or 
the Hamiltonian of the harmonic oscillator.
Thus, $\tilde{J}^{\pm}$ are now ladder operators of (what is left of) the 
$SU(2)$ Casimir. 
This means that they take us from one $SU(2)$ representation to another 
along an infinite 
``equal $\tilde{J}^3$ line." So now instead of a ``symmetry algebra" we have a 
``spectrum generating algebra" since the operators do not commute with the 
Hamiltonian. But we do not lose states. All the states are there but under 
different ``names" (quantum numbers).
Note that all the states with finite $J^3$ quantum numbers end up in 
the zero mode sector of the contracted algebra due to \mmm.
 
We believe a similar phenomenon occurs in the contraction of the $SO(2,4)\times 
SO(6)$ to the pp-wave. The pp-wave algebra has the same number of generators and 
the same number of supersymmetries as the original $AdS_5 \times S^5$. Some of 
the symmetries do not commute with the lightcone Hamiltonian but seen as a 
spectrum generating algebra it has to generate the full Hilbert space. It simply 
organizes it differently. The ``missing" states in the pp-wave with respect to 
$AdS_5 \times S^5$ should map to the zero mode sector $p_-=0$ since they did not 
scale fast enough to ``keep up" with the diverging denominator in \bmndic.

\subsec{Matrices and quivers.}

We end this section by pointing out an analogy between a class of operators 
discussed in \mrv\ and the matrices of M(atrix) theory.
The size of the matrices in M(atrix) theory with $N$ units of longitudinal 
momentum \susn\ is $N\times N$. In our case the $q$ units of longitudinal 
momentum \nulmpp\ relate to winding around a quiver diagram \mrv. 
The $|q=1,m=0\rangle$ states introduced in \mrv\ wind once around 
the quiver diagram (see equation (27) and figure 2. in \mrv)
\eqn\dizi{|q=1,m=0\rangle\sim {\rm Tr}(A_1A_2\dots A_M),} with $A_i$ the 
bi-fundamental 
fields in the $(N,\bar{N})$ of $SU(N)_i\times SU(N)_{i+1}$.
The general $|q,m=0\rangle$ state winds $q$ times around the quiver.
\eqn\diziq{|q,m=0\rangle\sim \underbrace{Tr\left( A_1A_2\dots A_MA_1A_2\dots 
A_M\dots\dots\dots A_1A_2\dots A_M \right)}_{q \quad times}.}
It is suggestive to relate \dizi\ with $1\times 1$  matrices in M(atrix) theory 
and \diziq\ with $q\times q$ matrices.
Perhaps the analogue of a ``block diagonal" matrix is a ``multi-trace" operator 
of the general form (e.g. in the case of minimal blocks)
\eqn\diziqdt{|q,m=0\rangle\sim \underbrace{Tr(A_1A_2\dots A_M)Tr(A_1A_2\dots 
A_M)\dots\dots\dots Tr(A_1A_2\dots A_M)}_{q \quad times}.} 
It would be very interesting to follow this suggestion further by interpreting 
the mixing between single and multi trace operators of this form as  
interactions between partons.
It is also interesting to understand the role of the winding modes $m\neq 0$ 
from this point of view.

\newsec{Relation to Maldacena's limit.}

The Penrose limit blows up the vicinity of a null geodesic to become the whole 
space. This is very reminiscent of the ``near horizon limit" (NHL) of 
Maldacena. Note that the horizon is also a null hypersurface. Indeed the NHL is 
exactly analogous to the Penrose limit where instead of the null geodesic there 
is the null hypersurface constituting the horizon. The scaling introduced 
by Maldacena, namely
\eqn\adscft{\alpha'\go 0 \quad, \quad {r \over \alpha'}=r m_s^2 \sim 
fixed}
is very reminiscent of the one investigated in this paper 
\penscale\sscale\genscale. In Maldacena's limit the 
radial coordinate transverse to the D3 branes plays the role of the longitudinal 
$x^-$ coordinate. 
This radial coordinate is transverse to the null surface just as in the Penrose 
limit the $x^-$ coordinate is transverse to the $x^+$ coordinate parameterizing 
the null geodesic. In both cases this direction is rescaled like the second 
power of the fundamental mass scale. However, there are also differences, e.g. 
the clean ``decoupling" property of Maldacena's limit may not occur in this 
case.  Also, in Maldacena's case there is no compactification involved.
This is analogous to BFSS and to BMN. Maldacena's limit seems to be related to a 
further generalization of DLCQ in the infinite longitudinal momentum limit.
We suspect that both procedures are special cases of a general rule (see also 
\refs{\aki,\joe}).

\newsec{Summary.}

In this paper we argued that string/M theory on the background of the Penrose 
limit of a spacetime is a generalization of the DLCQ procedure introduced by 
Seiberg and Sen.  We analyzed the case of type IIB strings on the maximally 
supersymmetric pp-wave which is the Penrose limit of $AdS_5 \times S^5$ and 
argued that it is analogous to the BFSS M(atrix) theory. The Penrose limit of an 
appropriate orbifold space $AdS_5 \times S^5/Z_M$ was understood as the analogue 
of Susskind's finite $N$ M(atrix) theory. We used this perspective to explain 
some of the features of the BMN theory.
We feel this gives better understanding of the nature of the BMN duality.

\centerline{\bf Acknowledgments} 

I would like to thank Emiliano Imeroni, Barak Kol, Kostas Skenderis and Jelper 
Striet for discussions and suggestions. 
I am very pleased to thank Riccardo Argurio for many helpful comments on the 
manuscript.   
Special thanks go to Jan de Boer for many discussions, suggestions and 
sharp questions as well as for comments on the manuscript.
This work is supported by a Clore fellowship.

\listrefs   
   
\end